\documentclass[11pt]{article}

\usepackage[lmargin=1.25in,rmargin=1.0in,tmargin=1.0in,bmargin=1.0in]{geometry}

\usepackage{epsfig}

\usepackage{sidecap}

\usepackage{indentfirst}

\usepackage{color}
\usepackage[small]{titlesec}
\titlespacing{\section}{0pt}{1em}{0.25em}

\newenvironment{myindentpar}[1]%
{\begin{list}{}%
         {\setlength{\leftmargin}{#1}}%
         \item[]%
}
{\end{list}}

\begin{document}

\noindent {\Large \bf MAESTRO, CASTRO, and SEDONA --- 
                      Petascale Codes for Astrophysical Applications}

\begin{myindentpar}{1in}
{\large \bf A. Almgren\footnote{Lawrence Berkeley National Laboratory}
        J. Bell\footnotemark[1], 
        D. Kasen\footnote{UC Santa Cruz}
        M. Lijewski\footnotemark[1], 
        A. Nonaka\footnotemark[1],
        P. Nugent\footnotemark[1],
        C. Rendleman\footnotemark[1],
        R. Thomas\footnotemark[1],
        M. Zingale\footnote{Stony Brook University}}
\end{myindentpar}

%\date{\today}

%\maketitle

\begin{myindentpar}{1in} % abstract to be indented 1 in
Performing high-resolution, high-fidelity, three-dimensional simulations of 
Type Ia supernovae (SNe Ia) requires not only algorithms that accurately
represent the correct physics, but also codes that effectively harness the 
resources of the most powerful supercomputers.  We are developing a suite 
of codes that provide the capability to perform end-to-end simulations of 
SNe Ia, from the early convective phase leading up to ignition to the explosion 
phase in which deflagration/detonation waves explode the star to the 
computation of the light curves resulting from the explosion.  In this paper 
we discuss these codes with an emphasis on the techniques needed to scale them 
to petascale architectures.  We also demonstrate our ability to map data from a 
low Mach number formulation to a compressible solver.
\end{myindentpar}

\section{Introduction}
We present a suite of codes for studying astrophysical phenomena
whose target is the end-to-end simulation of a Type Ia supernova (SN Ia)
at the petascale.  Each code is designed to perform optimally for a particular flow regime.
For the early convective phase of a carbon/oxygen white dwarf leading up to ignition,
we use MAESTRO \cite{MAESTRO}, a hydrodynamics code based on
a low Mach number approach that allows long-time integration of highly subsonic flow.  
The time step in MAESTRO is controlled by the fluid velocity instead of the sound speed, 
allowing a much larger time step than would be taken with a compressible code.
Once the star ignites and the fluid begins to travel at speeds 
no longer small relative to the speed of sound, the low Mach number assumption is invalid 
and the fully compressible equations must be solved to simulate the final seconds 
of stellar evolution before the explosion.
We simulate the explosion phase of SNe Ia with CASTRO \cite{CASTRO}, 
a fully compressible hydrodynamics code.  
Finally, SEDONA \cite{SEDONA},
a multidimensional, time-dependent, multi-wavelength radiation transport code,
is used to calculate the light curves and spectra from the resulting ejecta, enabling
direct comparison between computational results and observation.
All three codes have been designed to harness the resources of the
most powerful supercomputers available, and scale well to 100k-200k cores.

\begin{figure}
\centering
\includegraphics[width=0.40\textwidth]{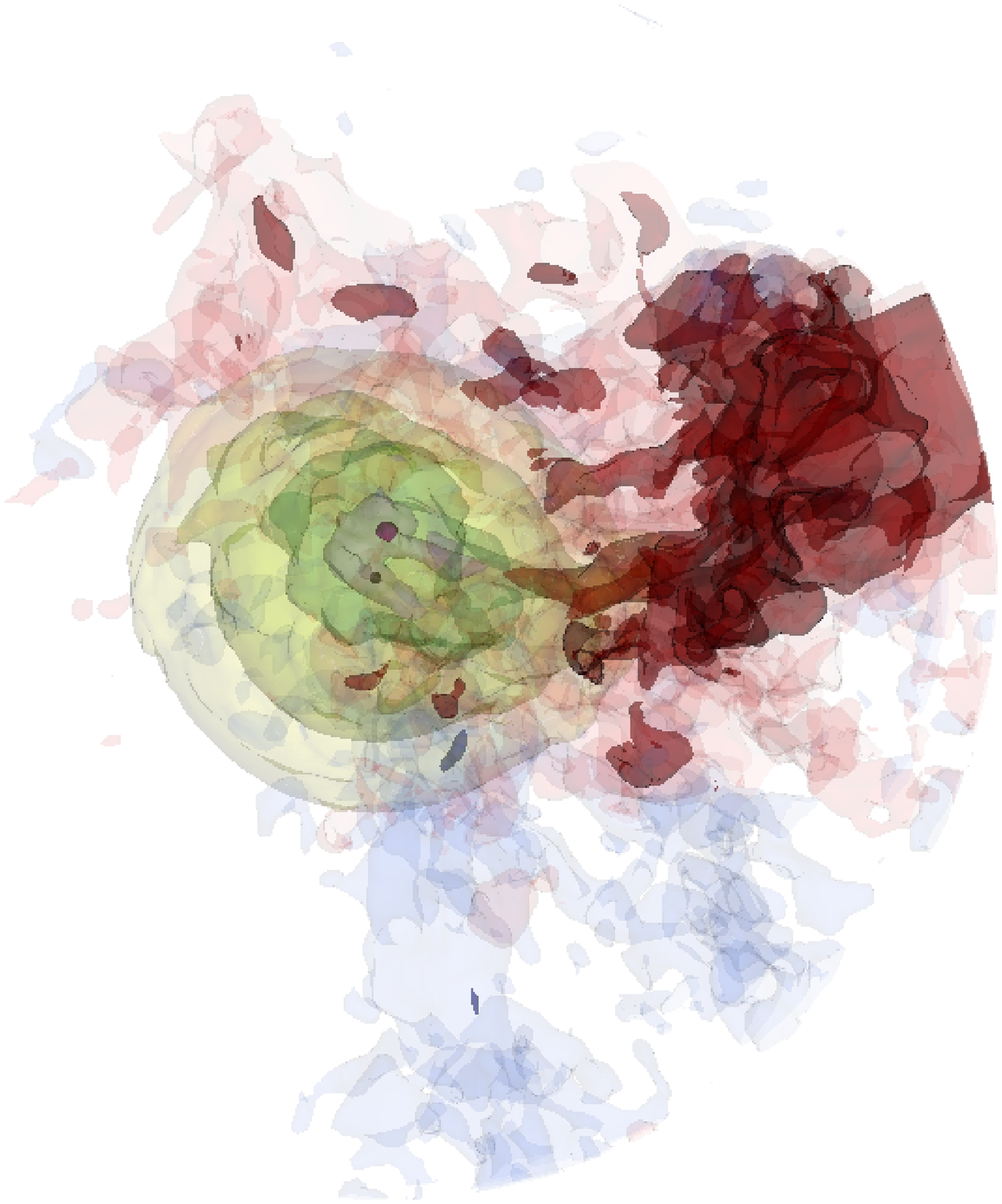}
\qquad\qquad
\includegraphics[width=0.40\textwidth]{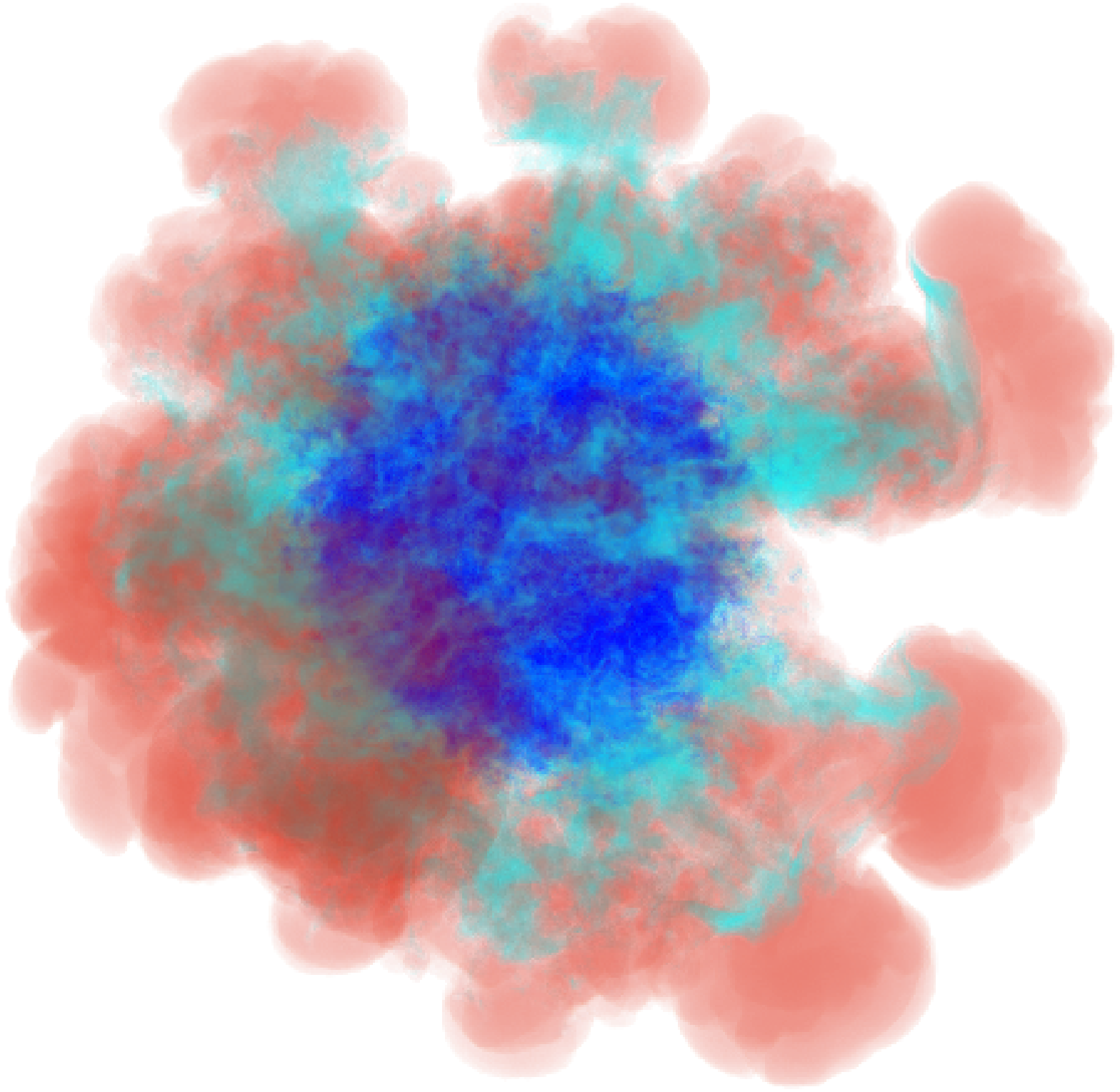}
\caption{(Left) MAESTRO simulation of convection in a white dwarf preceding a SN Ia.
Shown are contours of radial velocity (red=outward, blue=inward) and nuclear energy 
generation.  This simulation was performed using Jaguar at OLCF with 
an effective $768^3$ resolution and used approximately 1 million CPU-hours.
(Right) CASTRO simulation of nucleosynthesis during the explosion phase 
of a SN Ia.  Shown are the nuclear burning products (orange=iron, 
light blue=silicon and calcium, dark blue=helium).  This simulation was 
performed by Haitao Ma at UC Santa Cruz using Franklin at NERSC with an 
effective $8192^3$ resolution and used approximately 2 million CPU-hours.}
\label{fig:pretty_pics}
\end{figure}

MAESTRO and CASTRO use structured grids with adaptive mesh refinement (AMR);
SEDONA uses an implicit Monte Carlo approach.  A time step in CASTRO
requires the fully explicit advance of a hyperbolic system of conservation
laws, as well as the computation of self-gravity.
A time step in MAESTRO is composed of explicit advection as well the 
solution of a variable-coefficient Poisson equation 
that follows from the velocity constraint resulting from the low Mach number 
approximation.  A time step in each code also involves evaluations of the
equation of state as well as computation of any reactions.
In addition to simulations of SNe Ia (see Figure \ref{fig:pretty_pics}), 
CASTRO is also being used to study core-collapse and pair-instability supernovae,  
and MAESTRO is being applied to
convection in massive stars, X-ray bursts, and classical novae.

%%%%%%%%%%%%%%%%%%%%%%%%%%%%%%%%%
%%%%%%%%%%%%%%%%%%%%%%%%%%%%%%%%%

\section{Software Infrastructure}
MAESTRO and CASTRO are implemented using the BoxLib framework developed 
in the Center for Computational Sciences and Engineering at LBNL.
BoxLib is a hybrid C++ / Fortran90
software system that provides support for the development of parallel
structured-grid AMR applications.  In BoxLib, the memory management, 
flow control, parallel communication, 
and I/O are abstracted from the physics-specific routines, thus enabling
many different applications to be built on the same software framework.
SEDONA is implemented in a modern C++ framework that supports the
massively parallel Monte Carlo approach.

The fundamental parallel abstraction in BoxLib
is the MultiFAB, which holds the data on the union of 
disjoint rectangular grids at a level of refinement.
A MultiFAB is composed of FABs; each FAB is an array of data on a single grid. 
We use a coarse-grain parallelization strategy to distribute FABs to nodes, 
where the nodes communicate with each other using MPI.
We also use a fine-grain parallelization strategy in the physics-based modules
and the linear solvers, in which we use OpenMP to spawn a thread on
each core on a node.  Each thread operates on a portion of the associated FAB.
FABs at each level of refinement are distributed independently.

Each node contains meta-data that is needed to fully
specify the geometry and node assignments of the FABs.  
At a minimum, this requires the storage of an array of
boxes specifying the index space region for each AMR level of refinement.
The meta-data can thus be used to dynamically evaluate the necessary
communication patterns for sharing data between nodes for operations
such as filling data in ghost cells and synchronizing the solution at different levels of refinement.
Evaluating these communication patterns
requires computation of the intersections of the grids themselves with 
rectangular patches that represent grids with ghost cells.
A simple, brute force algorithm for doing so requires $\mathcal O(N^2)$ operations, 
where $N$ is the number of grids.  
This operation becomes expensive for problems with large numbers of grids, 
so we have implemented a hash sorting algorithm to reduce the cost.
Essentially, we subdivide the domain into multiple rectangular regions of index space, 
and sort the grids into these regions based on the lowest value in index space of each grid.
Each region is large enough that a grid based in one region extends no further than
the nearest neighbor regions.
We use the knowledge of which region each grid ``lives in '' to restrict our 
search for intersecting grids to only that region and its neighbors.
If $M$ is the number of regions covering the domain, this
reduces an $\mathcal O(N^2)$ operation to an $\mathcal O(N + N^2/M)$ operation.
In order to reduce the number of times the hash sort is called, we 
cache communication patterns that are most frequently used.

\section{Scaling Results}
\begin{figure}[h]
\includegraphics[width=0.5\textwidth]{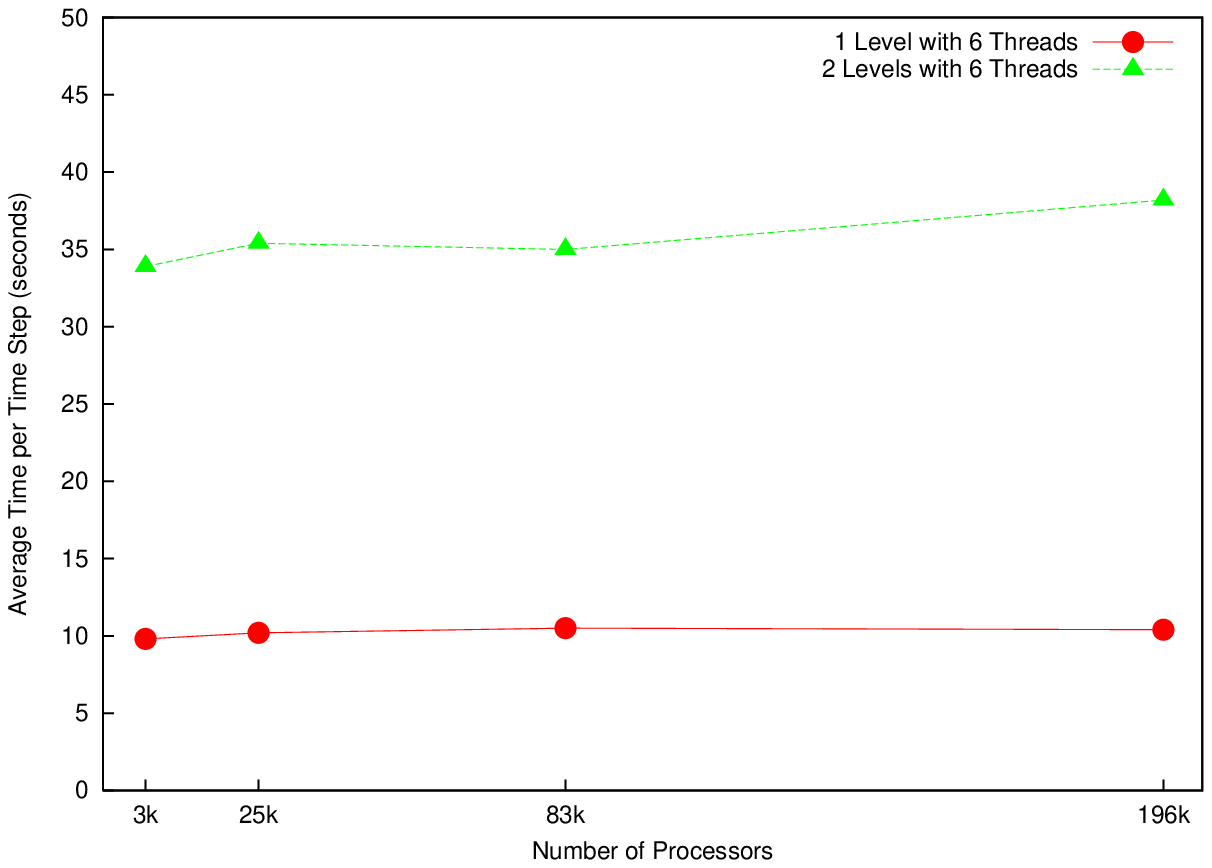}
\includegraphics[width=0.5\textwidth]{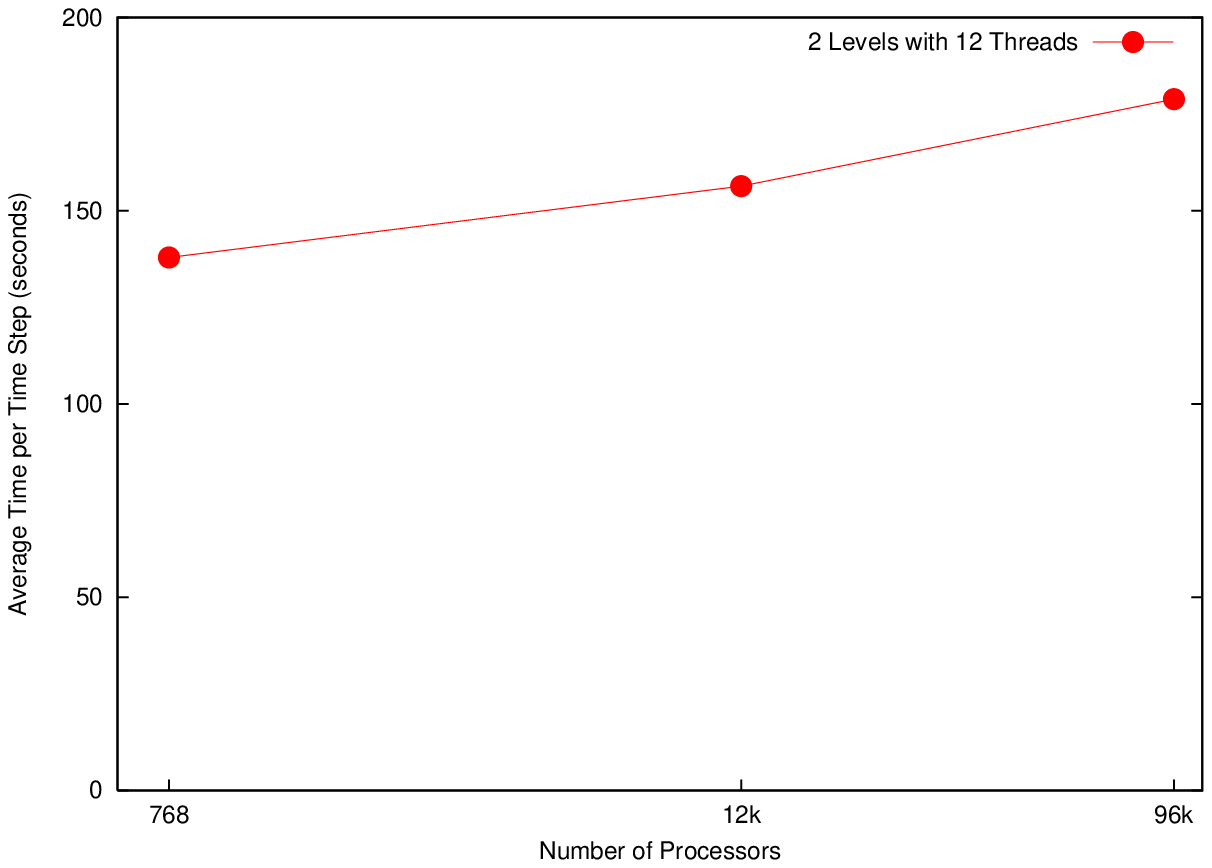}\\[3mm]
\includegraphics[width=0.5\textwidth]{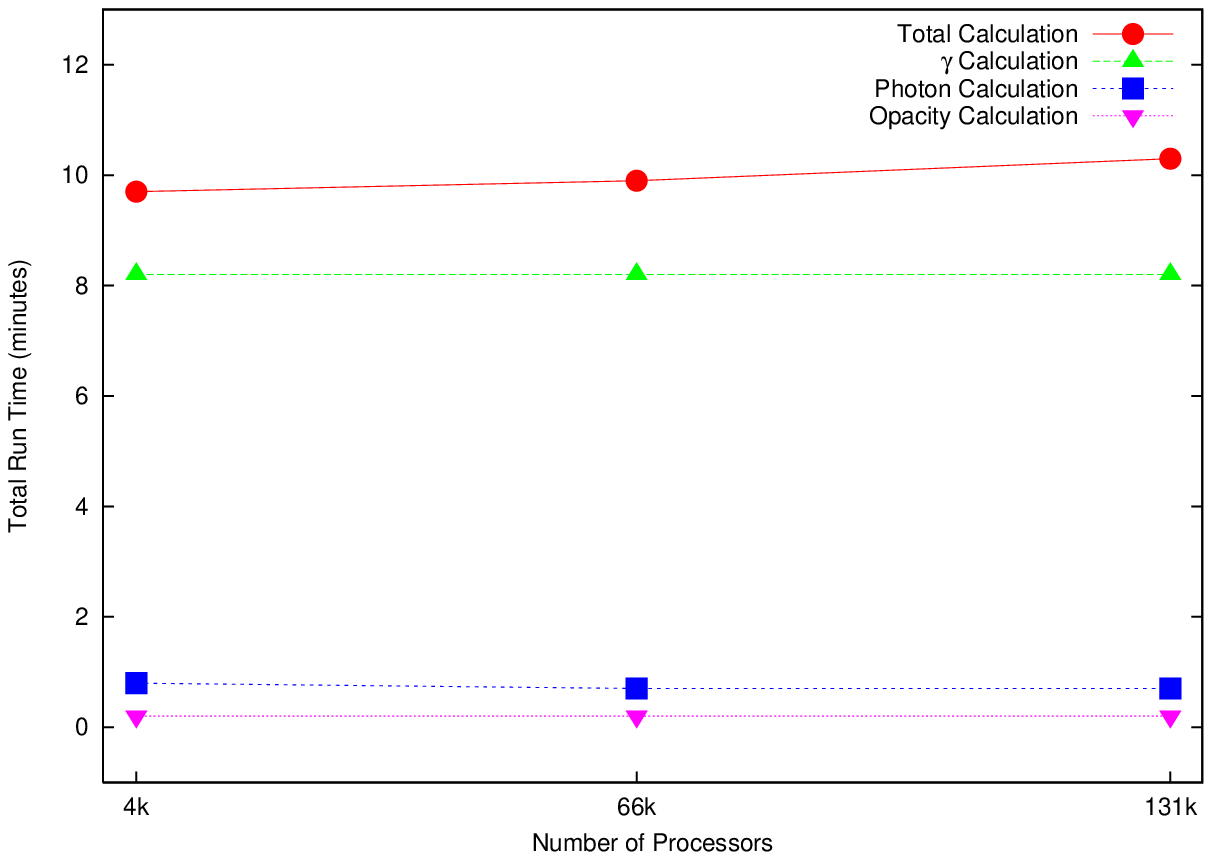}
\hspace{1em}
\begin{minipage}[b]{2.85in}
\caption{(Top-Left) CASTRO scaling results and (Top-Right) MAESTRO scaling results
on Jaguar at OLCF.  (Bottom) SEDONA scaling results using Intrepid at ANL.
Each test uses a weak scaling approach in which the number of cores 
increases by the same factor as the number of unknowns in the problem.
With perfect scaling the curves would be flat.} \ \\
\label{fig:scaling}
\end{minipage}
\end{figure}

We present scaling results demonstrating that our codes can efficiently run 
on the largest supercomputers (see Figure \ref{fig:scaling}).  We use a weak
scaling approach, in which the number of cores increases by the same factor
as the number of unknowns in the problem.  For the MAESTRO runs, we keep 
the one-dimensional radial base state fixed in time for this study; for the
CASTRO runs we use the monopole approximation for self-gravity.
In the MAESTRO and CASTRO tests, we simulate a full star on a 
three-dimensional grid. 
%and includes hydrodynamics, reactions, and self-gravity.
In the multilevel calculations the inner 12.5\% of the domain is refined.  
The results were obtained using Jaguar at 
OLCF, in which two hex-core sockets share memory on a node.  Thus we either
assign one MPI process per socket (in which case we spawn 6 threads), or one
MPI process per node (in which case we spawn 12 threads). In each case, a
single thread is assigned to a single core.  We note that CASTRO scaling 
behavior is relatively insensitive to using 6 or 12 threads.  MAESTRO has better
scaling performance when using 12 threads at a cost of additional thread overhead
time due to threading across different sockets.  The SEDONA scaling test
was performed using Intrepid at ANL using a pure-MPI approach,
and shows the parallel performance expected of a Monte Carlo method.

%%%%%%%%%%%%%%%%%%%%%%%%%%%%%%%%%
%%%%%%%%%%%%%%%%%%%%%%%%%%%%%%%%%

As shown in Figure \ref{fig:scaling}, CASTRO scales well for the single-level and
multilevel problems.  We can also determine the AMR overhead using this data.
Because of subcycling in time, a coarse time step consists 
of a single step on the coarse grid and two steps on the fine grid. Thus, we 
would expect that the time to advance the multilevel solution by one coarse time 
step would be a factor of three greater than the time to advance the single-level coarse
solution by one coarse time step, plus any additional overhead associated with AMR.
From the data in the figure we conclude that AMR introduces a modest overhead, ranging 
from approximately 15\% for the 4,000 core case to 18\% for the 196,000 core case.
By contrast, advancing a single-level calculation at the finer resolution by the 
same total time, i.e., two fine time steps, would require a factor of 16 more 
resources than advancing the coarse single-level solution.

The overall scaling behavior for MAESTRO is not as close to ideal as that of 
CASTRO due to the linear solves performed at each time step.  However, 
MAESTRO is able to take a much larger time step than CASTRO for flows in 
which the velocity is a fraction of the speed of sound, enabling the longer 
integration times needed to study convection.

\section{End-to-End Capability}

Performing an end-to-end simulation requires that a CASTRO simulation
be initialized with the correctly transformed data from a MAESTRO simulation, 
and that SEDONA be initialized with data from a CASTRO simulation.  
SEDONA takes as input the density, velocity and compositional structure 
of the material ejected in the explosion and synthesizes
emergent model spectra, light curves and polarization, which can then
be compared directly against observations.  This stage of the end-to-end
simulation capability is straightforward; the only remaining task is
to modify SEDONA to read the data from CASTRO's AMR hierarchy rather than from a uniform grid.
 
Initializing a CASTRO simulation with data from a MAESTRO simulation is analytically
more complicated due to the difference between the low Mach number approach and a 
fully compressible approach.  However, the fact that MAESTRO and CASTRO 
share a common software framework makes the implementation straightforward.
Here we demonstrate the successful mapping from MAESTRO to CASTRO for a 
two-dimensional test problem, that of an inflowing jet.  

%The output from CASTRO will then be synthesized
%by SEDONA, a time-dependent, multi-wavelength radiation transport code.  
%Thus, we will obtain emergent model spectra, light curves and 
%polarization, which can then be compared directly against observation, and will 
%complete our ``end to end'' study of SNe Ia.

%Since MAESTRO and CASTRO use the same software infrastructure, the checkpoint files
%are compatible with each other.  We demonstrate our ability to initialize a CASTRO
%simulation using a MAESTRO dataset using a two-dimensional inflow jet test problem.

\begin{SCfigure}
\includegraphics[width=0.60\textwidth]{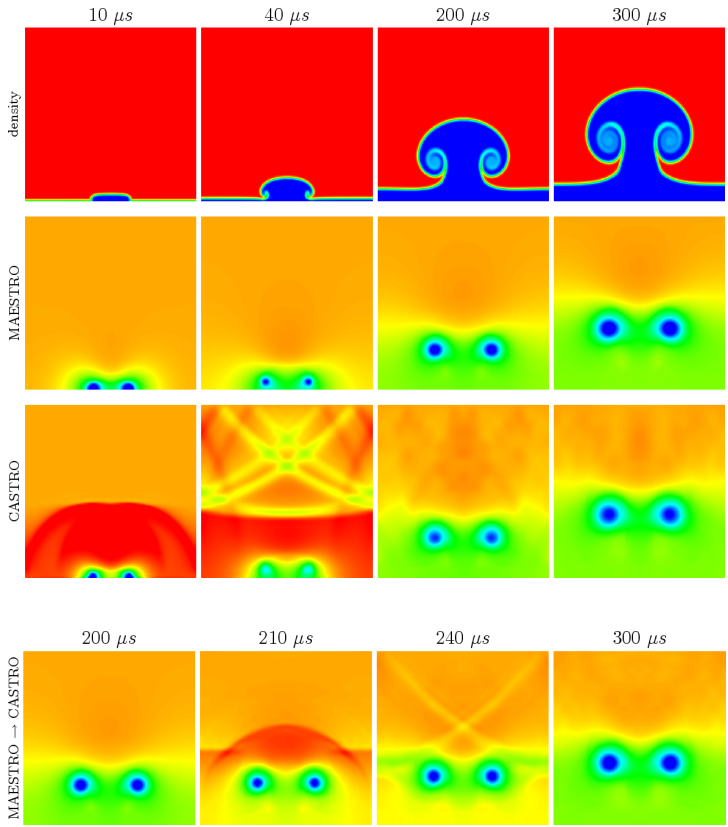}
\caption{Evolution of a low Mach number jet showing (top) density,
  (middle) MAESTRO pressure, (bottom) CASTRO pressure for the inflow
  jet problem at 4 different times in the evolution. The density plots
  are indistinguishable between the MAESTRO and CASTRO simulations.}
\label{fig:mach_jet_1}
\end{SCfigure}

\begin{SCfigure}
\includegraphics[width=0.60\textwidth]{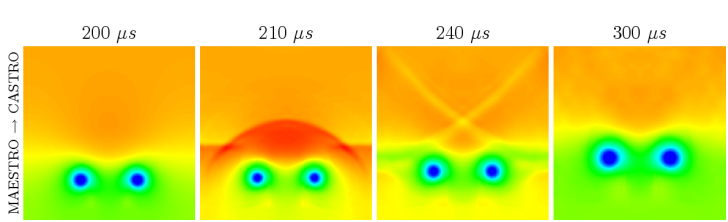}
\caption{Jet evolution using the MAESTRO dataset from $t=200~\mu\mathrm{s}$ (see 
Figure \ref{fig:mach_jet_1}) to initialize a CASTRO simulation.
Here, the time sequence corresponds to the last two columns
of Figure~\ref{fig:mach_jet_1}.}
\label{fig:mach_jet_2}
\end{SCfigure}

The computational domain is 1 cm on each side, and the pressure and density are set
to terrestrial conditions with zero initial velocity.  At the inflow face, we apply
a normal velocity with a maximum Mach number of 0.1, specifically,
\begin{equation}
v = c_s\left\{0.01 + 0.045[\tanh(100(x-0.4)) + \tanh(100(0.6-x))]\right\} ~ {\rm cm/s}.
\end{equation}
The inflow density is set to half of the initial value inside the domain.
In Figure \ref{fig:mach_jet_1}, we show the density and pressure fields computed
with MAESTRO and CASTRO to $t=300$ $\mu$s.  In the CASTRO simulation,
an acoustic wave is launched from the inflow boundary.  The acoustic signal
bounces around the domain until later times, when the solution has mostly
equilibrated.  In Figure \ref{fig:mach_jet_2}, we show the results from 
initializing a CASTRO simulation using the MAESTRO data from $t=200$ $\mu$s.
Shortly afterwards, the acoustic signal originating from the inflow boundary
has equilibrated, and the final-time data closely matches the simulations
in Figure \ref{fig:mach_jet_1}.
%%%%%%%%%%%%%%%%%%%%%%%%%%%%%%%%%
%%%%%%%%%%%%%%%%%%%%%%%%%%%%%%%%%
%%%%%%%%%%%%%%%%%%%%%%%%%%%%%%%%%
%%%%%%%%%%%%%%%%%%%%%%%%%%%%%%%%%

%\section{Summary}
%MAESTRO and CASTRO are mature astrophysical hydrodynamics codes, doing science, and
%ready for the petascale.  \MarginPar{need some sort of closing summary}

\section*{Acknowledgements}
We would like to thank Gunther Weber and Hank Childs of LBNL for their
help in using the VisIt visualization software. We thank 
Ken Chen, Candace Joggerst, Haitao Ma, and Jason Nordhaus 
for being patient early users of CASTRO, and Chris Malone for
his early work with MAESTRO. 
The work at LBNL was supported by
the SciDAC Program of the DOE Office of Mathematics, Information, and
Computational Sciences under the U.S. Department of Energy under
contract No.\ DE-AC02-05CH11231.  The work at Stony Brook was
supported by a DOE/Office of Nuclear Physics Outstanding Junior
Investigator award, grant No.\ DE-FG02-06ER41448, to Stony Brook.

\renewcommand{\refname}{\vskip -2em} 

%\bibliography{amr_code}

\end{document}